\RequirePackage{lineno} 
\documentclass[12pt,onecolumn,prc,showpacs,preprintnumber,amsmath,
floatfix,amssymb]{revtex4-1}

\usepackage{epsfig}
\usepackage{graphicx}
\usepackage{dcolumn}
\usepackage{bm}
\usepackage{multirow}           

\def\var{\mbox{\boldmath $\varepsilon$}}
\def\p{\mbox{\boldmath $p$}}
\def\q{\mbox{\boldmath $q$}}
\def\k{\mbox{\boldmath $k$}}

\begin{document}
\title{Flux-integrated inclusive and pionless cross sections for
  charged-current neutrino scattering off ${}^{40}\text{Ar}$ at energies
  available in the MicroBooNE experiment}
\author{A.~V.~Butkevich and S.~V.~Luchuk}
\affiliation{Institute for Nuclear Research,
Russian Academy of Sciences, Moscow 117312, Russia\\}
\date{\today}

\begin{abstract}
In this work, we analyze the flux-integrated inclusive and pionless
differential cross sections for neutrino scattering off argon. The cross
sections are calculated using the relativistic distorted-wave impulse
approximation (RDWIA), taking into account the contribution of the
two-particle-two-hole (2p-2h) meson-exchange currents (MEC). We find that the
measured single- and double-differential cross sections can be well described
within the experimental uncertainties using this approach. We also compare the
differential cross sections for pionless neutrino scattering on carbon and
argon, measured with the Booster Neutrino Beam flux in the MiniBooNE and
MicroBooNE experiments, to study nuclear effects in these nuclei.  
\end{abstract}
 \pacs{25.30.-c, 25.30.Bf, 25.30.Pt, 13.15.+g}

\maketitle

\section{Introduction}

A precise understanding of neutrino-nucleus interaction cross sections,
achieved at the percent level, is of paramount importance for neutrino
oscillation experiments~\cite{NOvA, T2K, MicroB1}. Such high accuracy is
critical not only for resolving the neutrino mass hierarchy and measuring the
CP-violating phase in the leptonic sector, but also for rigorously testing the
three-flavor neutrino framework. This requirement is particularly urgent for
next-generation long- and medium-baseline experiments, such as
SBN~\cite{SBN}, DUNE~\cite{DUNE}, Hyper-Kamiokande~\cite{HK}, and
JUNO~\cite{JUNO},
where systematic errors in cross-section models could become the dominant
limitation. Neutrino experiments extensively use scintillation and
water-Cherenkov detectors, providing abundant data on neutrino scattering
cross sections off carbon and oxygen nuclei. The experiments, including DUNE
and the SBN program, employ liquid argon time projection chambers (LArTPCs) as
detectors. Therefore, neutrino-argon cross section measurements are essential,
particularly in view of the relative scarcity of data on this nucleus.
 
Over the last decade, the MicroBooNE experiment has measured
inclusive and semi-inclusive neutrino-argon scattering cross sections using
the Booster Neutrino Beam (BNB) flux~\cite{BNB} at energies of
$0 < \var_{\nu} < 3$ GeV~\cite{Micro2, Micro3, Micro4, Micro5, Micro6}. In this
energy range,  charged-current (CC) quasielastic (QE) scattering and scattering
induced by two-body meson exchange current (MEC), resonance production (RES),
and deep-inelastic process (DIS) yield the dominant contribution to the neutrino
nucleus interaction.

The inclusive $\nu_{\mu}$ charged-current process, in which only the
outgoing muon is required to be reconstructed, comprises multiple interaction
mechanisms and is dominated by quasielastic scattering in the case of
MicroBooNE. Inclusive measurements are particularly
important as the clear signal definition allows a straightforward comparison
with theoretical models and other experiments. They are also the foundation
for studies of more complex event topologies involving detection of hadrons in
the final state. Inclusive predictions inherently integrate
over all final momenta of the initial interaction and give no information about
the subsequent evolution of the system. Thus, even if a model successfully
describes inclusive scattering, the final-state nucleon kinematics may be
inaccurate, causing it to be unable to describe semi-inclusive scattering data
and the details of the hadrons in the final state. A consistent theory should
be able to describe data for inclusive and semi-inclusive cross sections
covering the entirety of available phase space, as is needed for neutrino
oscillation experiments.

Recent years have seen a plethora of studies on neutrino-argon interactions
(see, e.g., Refs.~\cite{BAV1, Martini, Franco, Sob}).
Furthermore, official MicroBooNE publications have already provided
various comparisons with neutrino event generators. Similar analyses,
evaluating MicroBooNE and alternative data against diverse combinations of
generators, were also conducted in Refs.~\cite{Rocco, Macho, Dolan, Dytman}.
This activity is motivated by the
need to interpret data from the MicroBooNE experiment and to prepare for the
DUNE project. It has been shown that existing models and the event generators
that employ them are currently unable to describe data with a high level of
accuracy. This creates a need
for a diverse set of detailed neutrino-nucleus cross-section measurements
within the SBN program toe benchmark refinements to models and event
generators. Such measurements will stimulate improvements in theoretical
modeling, ultimately enhancing the sensitivity of future neutrino
experiments in a variety of ways.

We calculate the flux-integrated inclusive and pionless single- and
double-differential cross sections for the Booster Neutrino Beam flux,
comparing our results with MicroBooNE data~\cite{Micro4, Micro5, Micro6}. At
the mean flux energy of approximately 0.8~GeV,
the measured cross sections are dominated by CCQE and meson-exchange current
(MEC) (i.e., CCQE-like) processes. The theoretical framework additionally
accounts for pion production via resonance excitation. Specifically, the joint
calculation of CCQE and MEC contributions to nuclear lepton scattering
uses the relativistic distorted-wave impulse approximation (RDWIA) for the
quasi-elastic channel, and the superscaling approach (SuSA) combined with MECs
in the two-particle-two-hole sector (the SuSA-MEC approach) for the
meson-exchange current contribution~\cite{BAV2}.

The RDWIA, initially designed for the description of exclusive reactions\cite
{Pick, Udias1, Kelly1} and later adopted for neutrino reactions, has been
successfully tested against inclusive (e,e') data~\cite{BAV3}. This
macroscopic and unfactorized RDWIA takes into account the nuclear shell
structure and the
final-state interaction of the knocked-out nucleon with the residual nucleus.
In our approach, the effects of short-range nucleon-nucleon ($NN$) correlations
leading to the appearance of high-momentum and high-energy distributions in the
target are estimated~\cite{BAV3}. The flux-integrated semi-exclusive cross
sections for CCQE neutrino scattering off argon were calculated within the
RDWIA and verified against MicroBooNE data~\cite{Micro2}. The semi-exclusive
flux-integrated reduced cross sections for neutrino CCQE scattering on carbon,
oxygen, and argon were analyzed within this approach in Ref.~\cite{BAV5}.
As functions of the missing nucleon momentum, these cross sections exhibit
strong similarities to those of electron scattering and show good agreement
with the corresponding electron-scattering data for all three nuclei.
In Refs.~\cite{BAV6,BAV7}, the reduced cross sections for scattering on carbon
and oxygen were also calculated within the framework of the CCQE
scattering models used in the GENIE generator.

The SuSAv2 model exploits similarities between electron and neutrino
interactions to guide the description of weak scattering
processes~\cite{Meg1, Meg2}. In Ref.~\cite{BAV8}, a fit of the
RDWIA+MEC approach to the MiniBooNE neutrino CCQE-like (pionless, $CC0\pi$) data
was performed, and the best-fit value of the nucleon axial mass was obtained.
With this value of $M_A=1.2$~GeV the calculated single- and double-differential
cross sections are in good agreement with the MiniBooNE data.
The objective of this study is twofold. First, the flux-integrated inclusive
and pionless single- and double-differential cross sections are evaluated
within the RDWIA+MEC approach and subsequently validated against MicroBooNE
data. Second, a comparative analysis is performed on the differential cross
sections as functions of muon energy for neutrino interactions on argon and
carbon nuclei, as measured by the MicroBooNE and MiniBooNE~\cite{MiniB}
collaborations, respectively. This evaluation is directly related to the
scientific program of the upcoming SAND~\cite{SAND} detector at DUNE. By
benchmarking these experimental data, the SAND initiative seeks to refine
multi-target cross-section models for both carbon and argon within a unified
detector framework, thereby minimizing systematic uncertainties associated with
nuclear medium effects.

This article is organized as follows. In Sec.~II, we briefly present the
RDWIA+MEC approach and the model configuration employed in the neutrino event
generator GENIE version 3 simulation framework for the calculation of resonance
 and deep-inelastic neutrino scattering off nuclei. The results are presented
 and discussed in Sec.~III. The conclusions are summarized in Sec.~IV.


\section{Formalism of quasi-elastic scattering, RDWIA, }

In this work, we analyze the inclusive charged-current quasielastic
neutrino scattering off nuclei
\begin{equation}\label{qe:incl} 
\nu_{\mu}(k_i) + A(p_a) \rightarrow \mu(k_f) + X 
\end{equation} 
in the one $W$-boson exchange approximation.
In this expression, $k_i=(\varepsilon_i,\k_i)$ and $k_f=(\varepsilon_f,\k_f)$
are the initial
and final lepton momenta, $p_A=(\varepsilon_A,\p_A)$ is the momentum of the
target, $q=(\omega,\q)$ is the momentum transfer carried by 
the virtual $W$ boson, and $Q^2=-q^2=\q^2-\omega^2$ is the $W$-boson virtuality.

\subsection{CCQE-like quasielastic lepton-nucleus cross sections} 

In the inclusive processe described in Eq.~(\ref{qe:incl}),  
only the final-state lepton is detected. Consequently, the corresponding
differential cross section is expressed as follows:
\begin{eqnarray}
\frac{d^3\sigma}{d\varepsilon_f d\Omega_f} =
\frac{G^2\cos^2\theta_c}{2(2\pi)^2}\frac{|\mathbf{k}_f|}{\varepsilon_i}
L_{\mu \nu} W^{\mu \nu},
\label{CS}
\end{eqnarray}
where $\Omega_f=(\theta,\phi)$ denotes the solid angle of the scattered muon,
$G \simeq 1.16639 \times 10^{-11}$~MeV$^{-2}$ is the Fermi coupling constant,
and $\theta_C$ is the Cabibbo angle ($\cos \theta_C \approx 0.9749$). The
leptonic and weak charged-current hadronic tensors are represented by
$L_{\mu \nu}$ and $W^{\mu \nu}$, respectively. 

By decomposing the hadronic tensor into specific nuclear response functions,
the differential
cross section can be rewritten as follows:
\begin{eqnarray}
\frac{d^3\sigma}{d\varepsilon_f d\Omega_f} =
\frac{G^2\cos^2\theta_c}{(2\pi)^2} \varepsilon_f |\mathbf{k}_f| 
\left( v_0R_0 + v_TR_T + v_{zz}R_{zz} - v_{0z}R_{0z} - hv_{xy}R_{xy} \right).
\label{CSR}
\end{eqnarray}
Here, the coupling coefficients $v_k$ are kinematic factors depending on the
lepton variables, and their explicit analytical forms can be found in
Ref.~\cite{BAV3}.

The response functions are given in terms of components of the hadronic tensor
\begin{subequations}
\begin{align}
R_0 & = W^{00},\\
R_T & = W^{xx} + W^{yy},\\
R_{0z}&  = W^{0z} + W^{z0},\\
R_{zz} & = W^{zz}, \\                                             
R_{xy} & =  i\left(W^{xy}-W^{yx}\right) 
\end{align}
\label{4a-4e}
\end{subequations}
and depend either ($Q^2, \omega$) or ($|\q|,\omega$).   
All the nuclear structure information and final-state interaction effects are
contained in the weak CC nuclear tensor.
It is given by the bilinear products of the transition matrix 
elements of the nuclear CC operator $J^{cc}_{\mu}$ between the initial nuclear
state $|A\rangle$ and the final state $|X_f\rangle$ as 
\begin{eqnarray}
W_{\mu \nu } &=& \sum_f \langle X_f\vert                           
J^{(cc)}_{\mu}\vert A\rangle \langle A\vert
J^{(cc)\dagger}_{\nu}\vert X_f\rangle,              
\label{W}
\end{eqnarray}
where the sum runs over undetected states $X_f$. This equation~\ref{W} includes
all possible final states. Thus, the hadron tensor can be expanded as the sum of
the $1p-1h$ and $2p-2h$, plus additional channels, 
\begin{eqnarray}
W^{\mu \nu } &=& W^{\mu \nu}_{1p1h} + W^{\mu \nu}_{2p2h} + \cdots,      
\label{W_12}
\end{eqnarray}
where the $1p-1h$ channel yields the CCQE response functions and the $2p-2h$
hadronic tensor determines the $2p-2h$ MEC response functions. 
Consequently, each nuclear response function $R_i$ defined in Eqs.~(4a)--(4e)
can be decomposed into a linear combination of the quasielastic ($R_{i,QE}$)
and meson-exchange current ($R_{i,MEC}$) components:
\begin{eqnarray}
R_i &=& R_{i,QE} + R_{i,MEC}.                                      
\label{R_12}
\end{eqnarray}

Within the framework of the impulse approximation, we describe genuine CCQE
neutrino-nucleus scattering by assuming that the incident neutrino interacts
with a single nucleon. This nucleon is then ejected from the nucleus, whereas
the other (A-1) target nucleons remain passive spectators. Consequently, the
total nuclear current is constructed as a superposition of individual
single-nucleon currents. The single-nucleon charged current has a $V{-}A$
structure $J^{\mu} = J^{\mu}_V +J^{\mu}_A$. For the free-nucleon vertex function
$\Gamma^{\mu} = \Gamma^{\mu}_V + \Gamma^{\mu}_A$ we use the vector current vertex
function
$\Gamma^{\mu}_V = F_V(Q^2)\gamma^{\mu} + {i}F_M(Q^2)\sigma^{\mu\nu}q_{\nu}/2m$,
where $m$ is the nucleon mass,
$\sigma^{\mu \nu}=i[\gamma^{\mu},\gamma^{\nu}]/2$ and $F_V$ and $F_M$
 are the weak vector form factors. Under the conserved vector current (CVC)
 hypothesis, the weak vector form factors are directly coupled to the
 respective electromagnetic form factors of the proton and neutron. For the
 evaluation of these vector nucleon form factors, we adopt the
parameterization proposed in Ref.~\cite{MMD}. To account for the off-shell
nature of the bound nucleons, we adopt the de~Forest prescription~\cite{deFor}
and evaluate the off-shell vector current vertex $F^{\mu}_V$ within the Coulomb
gauge.

Figure~1(a) and (b) show the various parametrizations of the isovector Sachs
electric $G_E$ and magnetic $G_M$ form factors as functions of $Q^2$. 
These form factors are related to the Dirac $F_V$ and Pauli $F_M$ form factors
by
\begin{equation}
 G_E=F_V - \frac{Q^2}{4m^2}F_M ~~~~~~G_M=F_V+F_M.
\end{equation}
As shown, the differences between the MMD parametrization~\cite{MMD} of the
vector form factors and the newer ones from Kelly~\cite{Kelly2},
Zex18~\cite{Zex18}, and NME~\cite{NME} are negligible in the
region $Q^2 \le 2$ (GeV/c)${}^2$. Note that the parametrizations from
Refs.~\cite{Zex18, NME} are derived from recent lattice quantum chromodynamics
(LQCD) simulations.
\begin{figure*}
  \begin{center}
    \includegraphics[height=14cm, width=14cm]{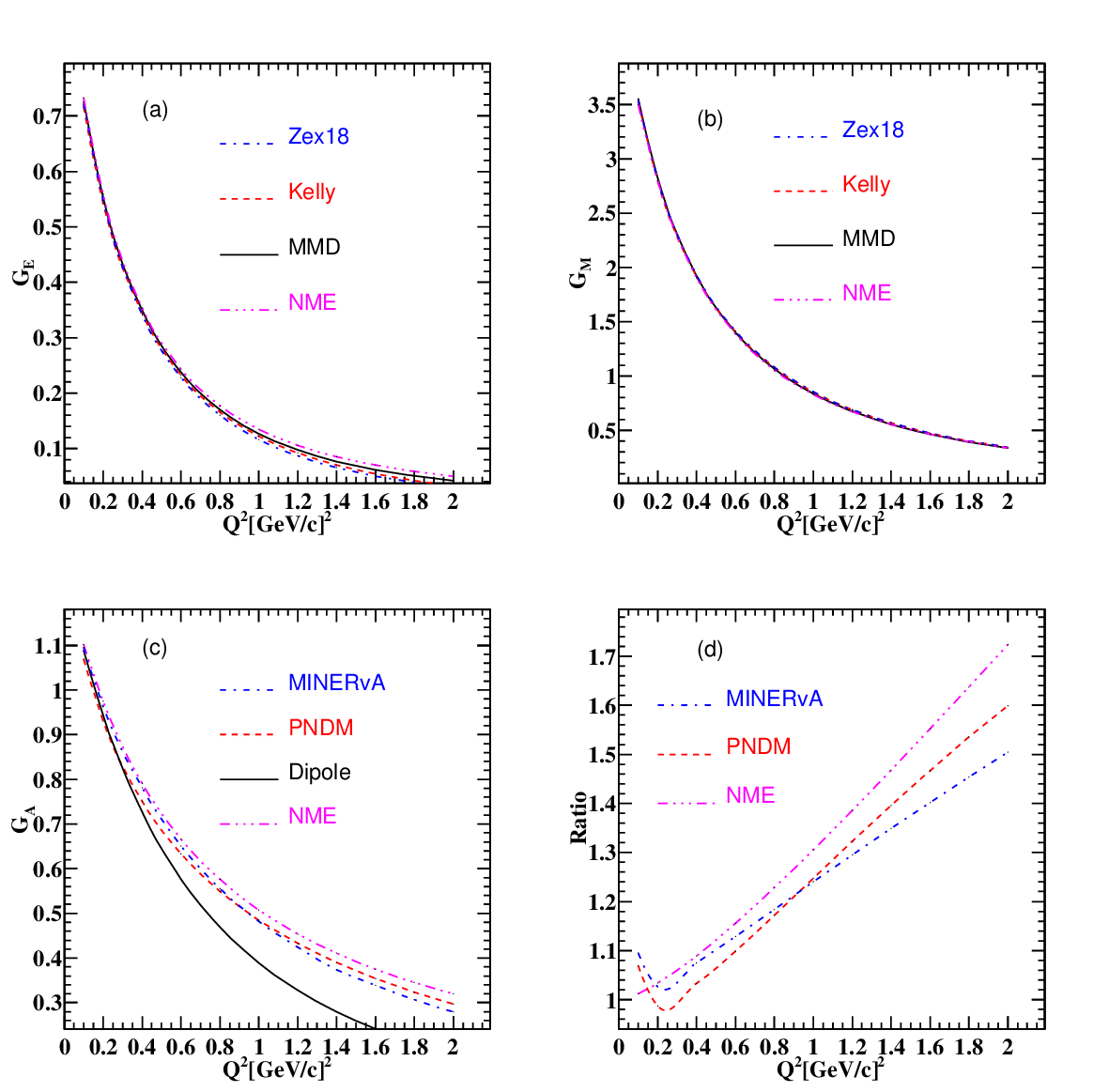}
  \end{center}
  \caption{\label{Fig1}
    Top panels: Sachs electric $G_E(Q^2)$ (a) and magnetic $G_M(Q^2)$
    (b) form factors from Refs.~\cite{MMD} (solid line), \cite{Kelly2}
    (dashed line), \cite{NME} (dotted line), and~\cite{Zex18} (dash-dotted line)
    plotted versus $Q^2$. Bottom panels: (c) axial form factor $G_A$
    from Refs.~\cite{PNDME} (dashed line), \cite{NME} (dotted line),
    \cite{Miner} (dash-dotted line), and the dipole ansatz with $M_A = 1.2$~GeV
    as a function of $Q^2$; (d) the ratio $G^{\text{LQCD}}_A/G^{\text{dipole}}_A$
    versus $Q^2$.
 }
\end{figure*}

The axial current vertex function can be written in terms of the axial
$F_A(Q^2)$ and pseudoscalar $F_P$ form factors:
\begin{equation}
\Gamma^{\mu}_A = F_A(Q^2)\gamma^{\mu}\gamma_5 + F_P(Q^2)q^{\mu}\gamma_5.      
\label{Eq8}
\end{equation}
The pseudoscalar form factor
is dominated by the pion-pole contribution. Invoking the hypothesis of a
partially conserved axial current (PCAC), this factor can be expressed via
 the Goldberger-Treiman relation in the low-$Q^2$ region ($Q^2\approx 0$). We
 assume that a similar relation is valid for high $Q^2$ as well.
To describe the $Q^2$ dependence of these form factors, a standard dipole
parameterization is employed:
\begin{equation}
F_A(Q^2)=\frac{F_A(0)}{(1+Q^2/M_A^2)^2},\quad                              
F_P(Q^2)=F_A(Q^2)F'_P(Q^2),
\label{Eq9}
\end{equation}
where $F'_P=2m^2/(m_{\pi}^2+Q^2), F_A(0)=1.2724$, $M_A$ is the axial mass, and
$m_\pi$ is the pion mass. Then, the axial current vertex function can be written
 as
\begin{eqnarray}
\label{Eq10}
\Gamma_A^{\mu} &=& F_A(Q^2)[\gamma^{\mu}\gamma_5 +F'_P(Q^2)q^{\mu}\gamma_5]  
\end{eqnarray}
and the axial-vector current can be factorized as
\begin{eqnarray}
\label{Eq11}
J_A = F_A(Q^2)J'_A(Q^2),                                    
\end{eqnarray}
where $J'_A = \gamma^{\mu}\gamma_5 +F'_P(Q^2)q^{\mu}\gamma_5$.

In this work, we adopt the value $M_A=1.2$~GeV, which was determined in
Ref.~\cite{BAV8} from an analysis of neutrino scattering data. Figure~1(c)
illustrates the dipole parametrization along with the $z$-expansion fits to $F_A$
obtained from the LQCD simulations in NME~\cite{NME}, PNDM~\cite{PNDME},
MINERvA~\cite{Miner}. Notably,
the dipole ansatz with this choice of $M_A$ predicts a steeper decline in the
axial form factor at higher $Q^2$ compared to the LQCD benchmarks. The ratio
$R=F^{LQCD}_A/F^{dipole}_A$ is plotted as a function of $Q^2$ in Fig.~1(d), where
$F^{LQCD}_A$ and $F^{dipole}_A$ denote the axial form factors calculated using the
LQCD $z$-expansion and the dipole parametrization, respectively. At $Q^2=1$
~(GeV/c)${}^2$, the discrepancy between the two approaches reaches
approximately 20\%.

\subsection{RDWIA model}

Within the framework of the RDWIA, the relativistic wave functions for bound
nucleon states are calculated in the independent particle shell model (IPSM) as
self-consistent solutions to the Dirac equation. This equation is derived using
a relativistic mean-field approach based on a Lagrangian that includes $\sigma,
 \omega$ and $\rho$ mesons (the $\sigma - \omega$ model)\cite{serot,horow}.

 These functions were calculated using the TIMORA code~\cite{horow} with
the normalization factors $S_{\alpha}$ relative to the full occupancy of the
IPSM orbital $\alpha$ of ${}^{40}$Ca. For ${}^{40}$Ca and ${}^{40}$Ar an average
factor is $\langle S \rangle \approx 87\%$. This estimation of depletion of hole
states follows from the RDWIA analysis of ${}^{40}$Ca$(e,e'p)$
data~\cite{BAV0}. The source of the reduction in the
$(e,e'p)$ spectroscopic factors with respect to the mean-field values is 
the short-range and tensor correlations in the ground state, which lead to the
appearance of the high-momentum and high-energy component in the nucleon
distribution in the target.

In the RDWIA approach, the effects of final state interactions (FSI) for the
outgoing nucleon are taken into account. The distorted-wave function
describing the knocked-out nucleon is determined by solving a Dirac equation
that includes a phenomenological relativistic optical potential. In this study,
we use the EDAD1 parametrization~\cite{Cooper} of the relativistic optical
potential tailored for carbon. The inclusive cross sections were computed with
this potential, keeping only its real part.
Experimental data from electron scattering measurements~\cite{BAV7, CaN1, CaN2,
  CaN3} indicate that this specific parametrization of the optical potential is
preferred over alternative parametrizations. 
The cross sections with FSI effects in the presence of  
short-range $NN$-correlations were calculated using the method proposed in
 Ref.~\cite{BAV1} with the nucleon high-momentum distribution from 
Ref.~\cite{Atti} renormalized to a value of 13\%.
Within this framework, the contribution from $NN$-correlated pairs is computed
in the impulse approximation, meaning that the virtual $W$-boson couples
exclusively to a single member of the $NN$-pair. This one-body mechanism results
in the emission of two nucleons, corresponding to a 2p-2h excitation.

\subsection{Non-CCQE contributions}

In this study, we compute the weak MEC response functions $R_{i,MEC}$
for neutrino-argon scattering by employing precise parametrizations derived
from exact MEC calculations~\cite{Amaro2}.
The elementary hadronic tensor is defined as a bilinear product of the matrix
elements of the two-body weak (vector and axial) MEC. Within this model, only
the one-pion exchange mechanism is taken into account. The two-body current
operator is obtained from the pion production amplitudes for the nucleon by
coupling a second nucleon to the emitted pion. The corresponding MEC operator
is expressed as a sum of seagull, pion-in-flight, pion-pole, and
delta-pole components. The $\Delta$ peak provides the dominant contribution to
the pion production cross section, and the MEC peak is located in the ``dip''
region between the QE and Delta peaks.
\begin{figure*}
  \begin{center}
    \includegraphics[height=17cm,width=19cm]{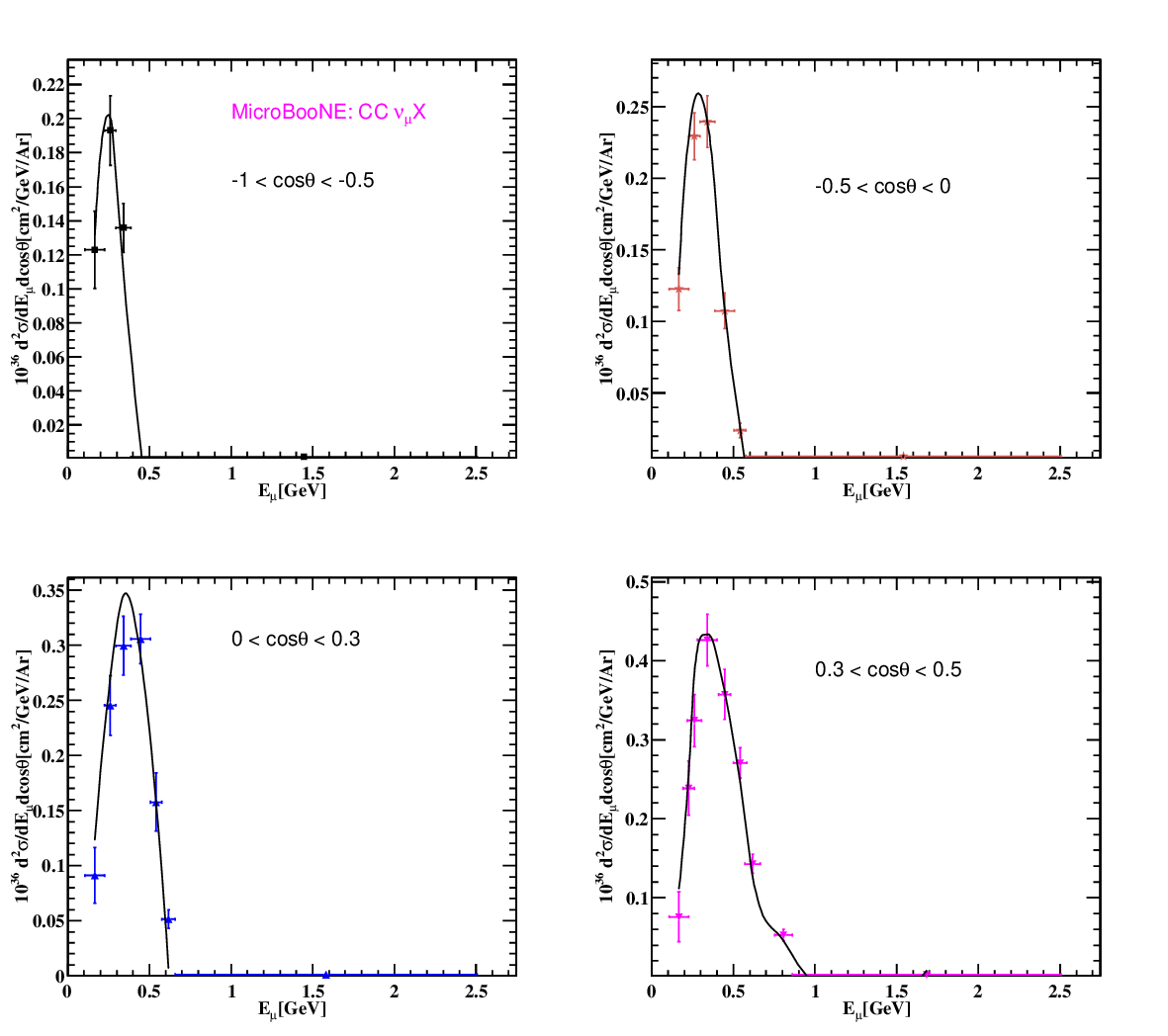}
  \end{center}
  \caption{\label{Fig2} Flux-integrated inclusive
    $d^2\sigma/d\var_{\mu} d\cos\theta$ cross section per argon nucleus for
    the $\nu_{\mu}\text{Ar}$ scattering as a function of $\cos\theta$ for the
    four muon scattering angle bins: $\cos\theta$=(-1 - -0.5), (-0.5 - 0),
    (0 - 0.3), and (0.3 - 0.5). 
The MicroBooNE data are shown as points.} 
\end{figure*}
Furthermore, the lepton-nucleus cross sections computed with these MEC
parametrizations have been successfully validated against world data for
electron scattering on nuclei~\cite{Amaro3, Megias2, BAV2}.
\begin{figure*}
  \begin{center}
    \includegraphics[height=17cm,width=19cm]{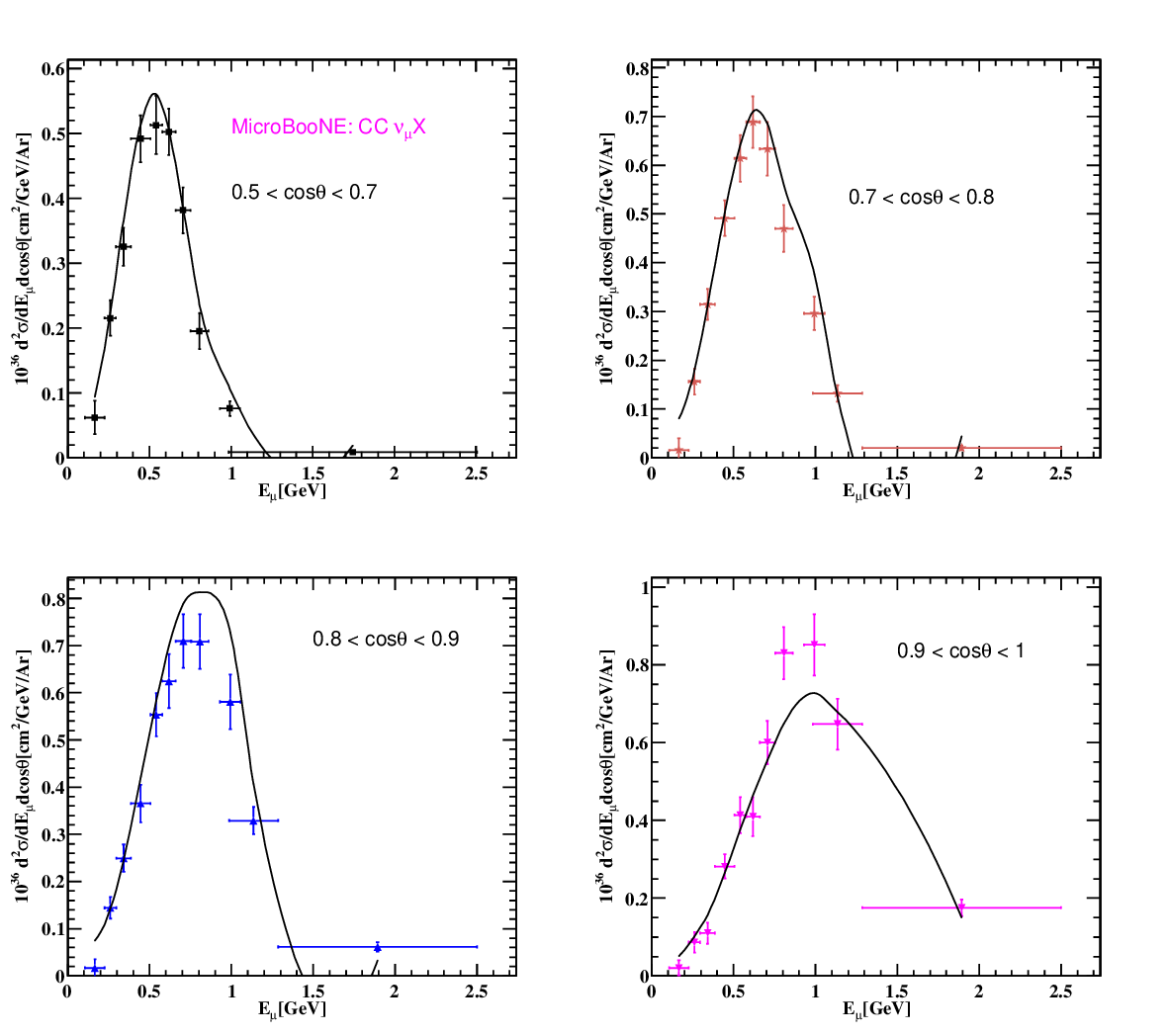}
  \end{center}
  \caption{\label{Fig3} Same as Fig.~\ref{Fig2}, but for the muon 
scattering angle bins: $\cos\theta$=(0.5 - 0.7), (0.7 - 0.8), (0.8 - 0.9), and 
(0.9 - 1).} 
\end{figure*}

The contributions of resonance production and deep-inelastic scattering to
the inclusive cross sections are evaluated using the GENIE neutrino event
generator (version 3.04.00)~\cite{Tena} with the G18-10a model configuration.
 This model set incorporates the Berger-Sehgal model~\cite{Berger} for resonant
channels and a scaled Bodek-Yang prescription~\cite{Bodek} for non-resonant
processes. Additionally, pion absorption is simulated via the hA2018
intranuclear cascade model~\cite{hA2018}.
\begin{figure*}
  \begin{center}
    \includegraphics[height=11cm, width=15cm]{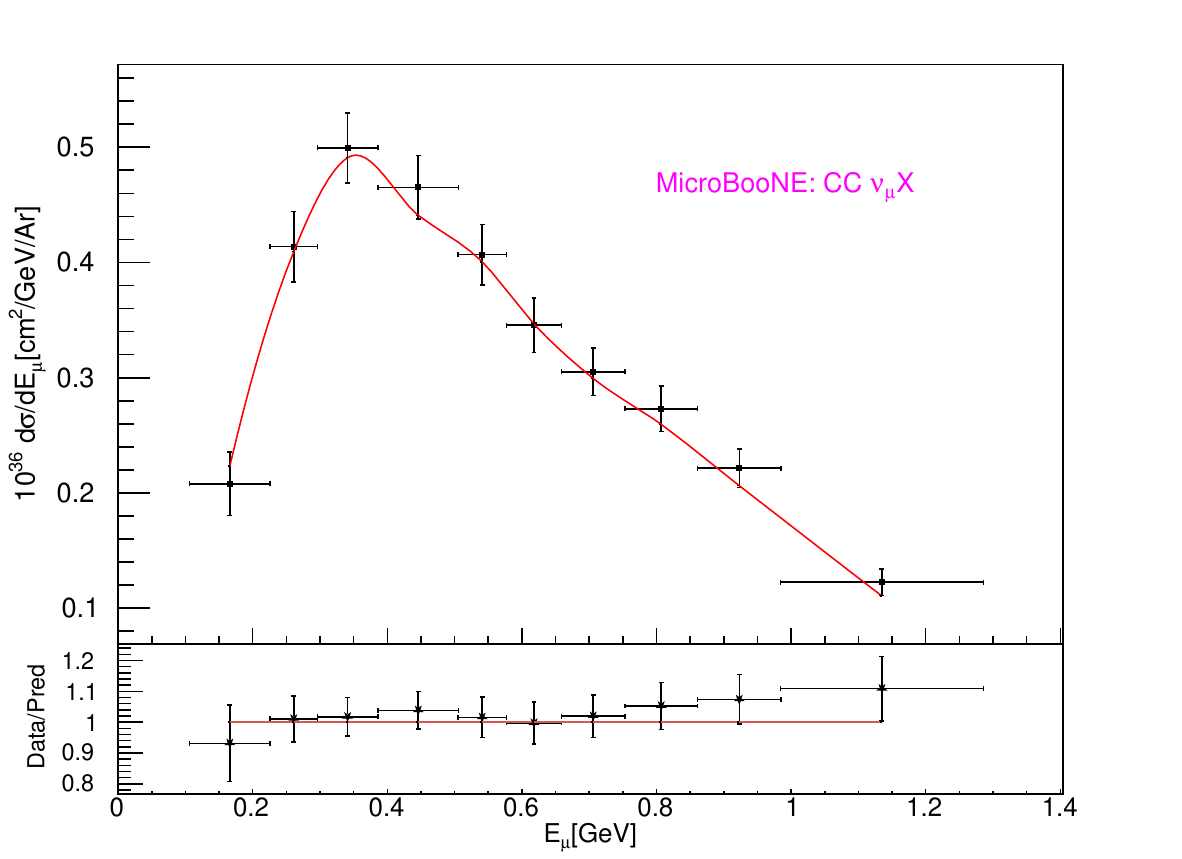}
  \end{center}
  \caption{\label{Fig4}
    Flux-integrated inclusive $d\sigma/dE_{\mu}$ cross section per argon
    nucleus as a function of the muon energy $E_{\mu}$. The bottom subpanels
    show the comparison between the data and the theoretical predictions.     
}
\end{figure*}
\section{Results and analysis}

The unfolded single- and double-differential cross sections per $\text{Ar}$
nucleus for inclusive and pionless $\nu_{\mu}\text{Ar}$ processes were extracted
in Refs.~\cite{Micro5} and~\cite{Micro6}, respectively. In this work, we
calculate these cross sections by integrating over the Booster Neutrino
Beamline flux~\cite{MiniB}. To compare our predictions with the data, we
smear them with the smearing matrix $A_c$ obtained in the unfolding
procedure.

\subsection{Flux-integrated inclusive differential cross sections for
  $\nu_{\mu}\text{Ar}$ scattering}

\begin{figure*}
  \begin{center}
    \includegraphics[height=12cm, width=18cm]{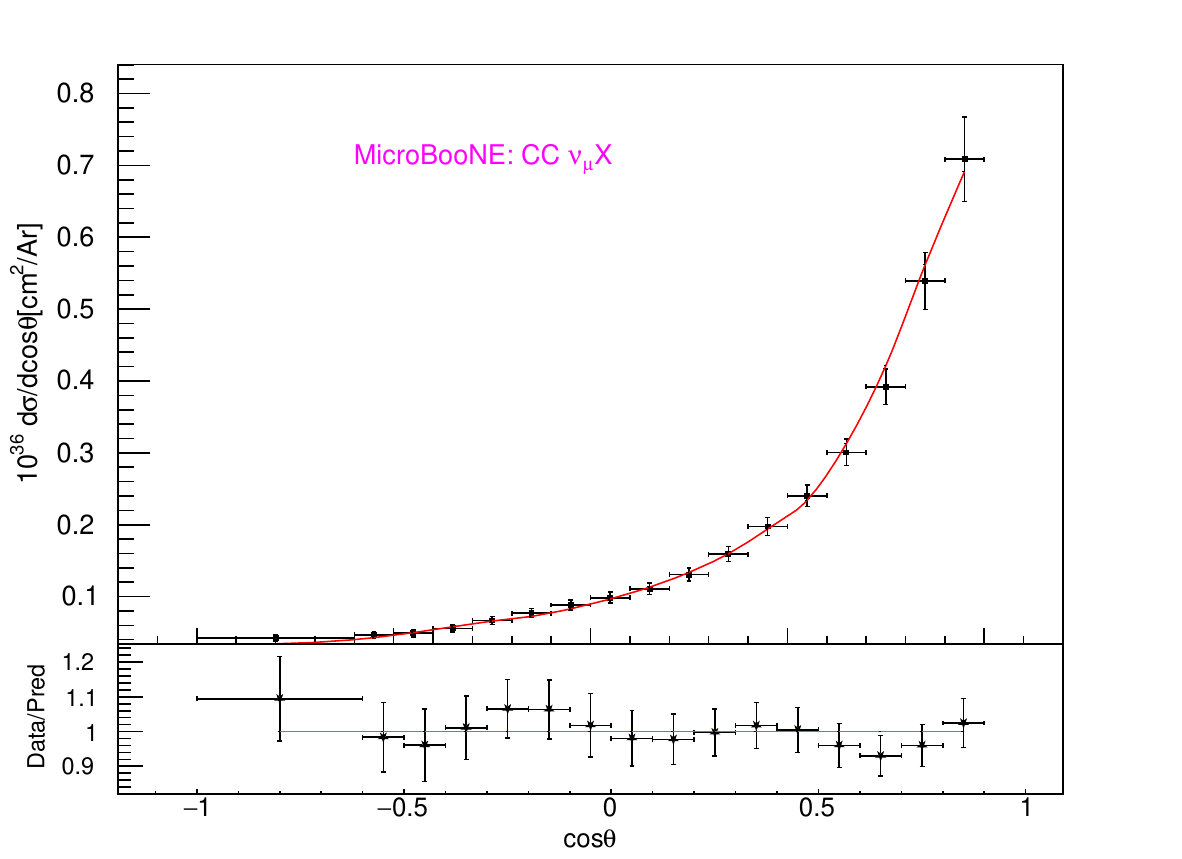}
  \end{center}
  \caption{\label{Fig5}
    Flux-integrated inclusive $d\sigma/d\cos\theta$ cross section per argon
    nucleus as a function of the muon scattering angle $\cos\theta$. The bottom
    subpanels show the comparison between the data and the theoretical
    prediction.     
}
\end{figure*}

We calculated the flux-integrated double differential cross section 
$d^2\sigma/d\var_{\mu} d\cos\theta$  using 
the $\var_{\mu}$ and $\cos\theta$-bins similar to Ref.~\cite{Micro5}

Figures~\ref{Fig2} and~\ref{Fig3} show measured flux-integrated 
$d^2\sigma/d\var_{\mu}d\cos\theta$ cross sections as functions of muon energy
for several bins of muon scattering  angle in the range
$-1\leq \cos\theta \leq 1$ as compared with calculations.
There is good agreement between the calculations and data within the error of
the experiment.
The experimental uncertainties, defined as the square root of the diagonal
elements of the extracted covariance matrix, exceed 10\%. However, the
calculated cross section at its maximum slightly overestimates
the measured data in the $0.8 < \cos \theta < 0.9$ range and underestimates it
in the $0.9 < \cos \theta < 1.0$ interval.
The flux-integrated single-differential cross sections, $d\sigma/d\var_{\mu}$ and
$d\sigma/d\cos\theta$ are presented in Figs.~\ref{Fig4} and~\ref{Fig5}.
Figure~\ref{Fig4} shows $d\sigma/d\var_{\mu}$ as a function of the muon energy,
  while figure~\ref{Fig5} displays $d\sigma/d\cos\theta$ as a function of the
  muon scattering angle.
\begin{figure*}
  \begin{center}
    \includegraphics[height=17cm,width=19cm]{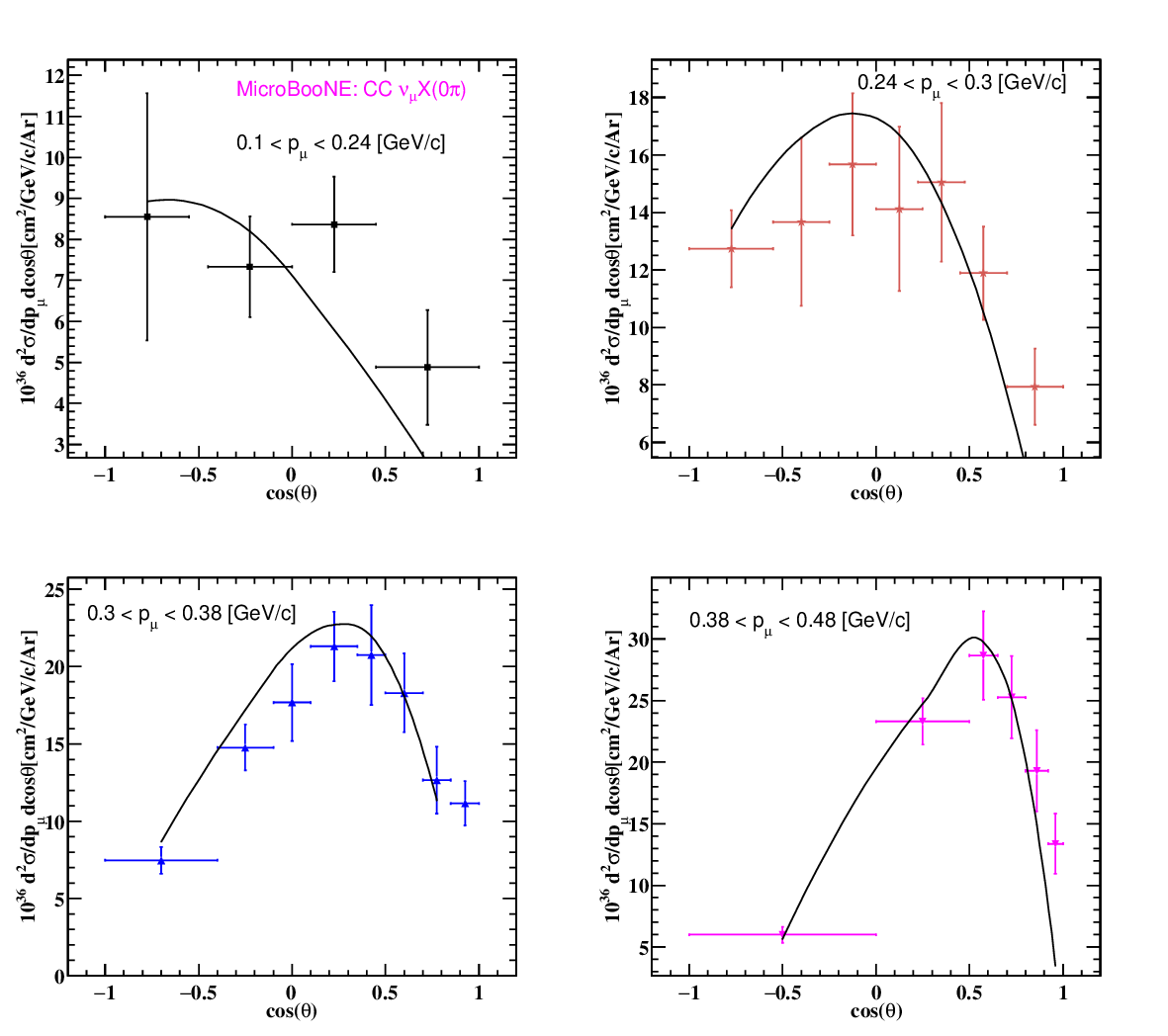}
  \end{center}
  \caption{\label{Fig6} Flux-integrated $d^2\sigma/dp_{\mu} d\cos\theta$
    cross section per argon nucleus for pionless
     $\nu_{\mu}\text{Ar}$ scattering as a function of $p_{\mu}$ for four
    muon momentum bins: $p_{\mu}$=(0.1 - 0.24), (0.24 - 0.3),
    (0.3 - 0.38), and (0.38 - 0.48)(GeV/c) . 
The MicroBooNE data are shown as points.} 
\end{figure*}
The calculated results are compared with the MicroBooNE data, and the
data-to-prediction ratios are also shown. The experimental uncertainties exceed
6\%. Good agreement is observed between the calculated and measured cross
sections within the experimental uncertainties.
\begin{figure*}
  \begin{center}
    \includegraphics[height=11cm,width=19cm]{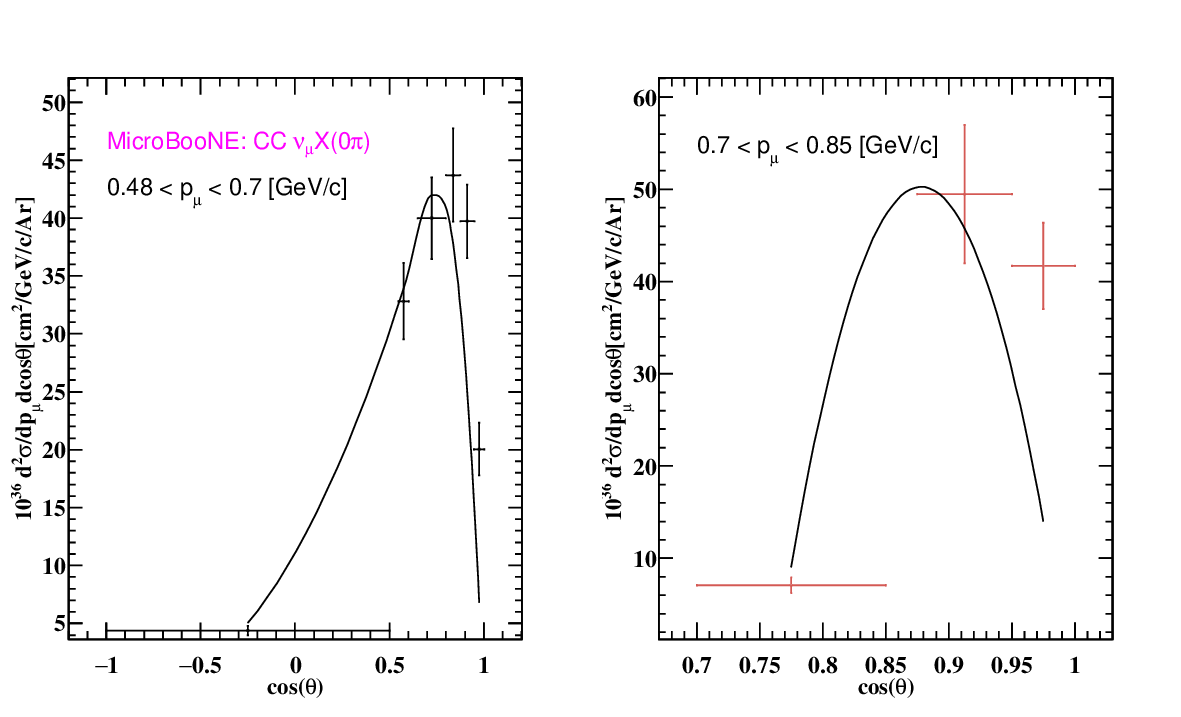}
  \end{center}
  \caption{\label{Fig7}Same as Fig.~\ref{Fig6}, but for muon momentum  
     bins: $p_{\mu}$=(0.48 - 0.7) and (0.7 - 0.85) (GeV/c).
The MicoBooNE data are shown as points.} 
\end{figure*}
However, at a muon energy of
$E_{\mu}>0.8$ GeV, a slight systematic underestimation of the measured cross
sections is observed. The contributions of the CC, MEC, RES, and DIS processes
to $d\sigma/d\var_{\mu}$ are approximately 50\%, 16\%, 26\%, and 8\%,
respectively. While the CCQE contribution is about 45\% in the peak region of
the cross section at a muon energy of 0.35~GeV, it reaches 65\% at energies
above 1.2~GeV.

\subsection{ Flux-integrated differential cross sections for pionless
  $\nu_{\mu}\text{Ar}$ scattering}

Using the G18-10a model configuration from the GENIE v3.4.0 neutrino event
generator, we estimated the probability of pion production with momentum
higher than 70~MeV/c in the final states of CCQE and MEC interactions at 
MicroBooNE energies. This probability was found to be less than 3\%. The
contribution of resonance interactions to pionless ($\text{CC}0\pi$) events at
these energies - i.e., events without pions with momentum higher than
70 MeV/c - is less than 10\%.
\begin{figure*}
  \begin{center}
    \includegraphics[height=11cm, width=15cm]{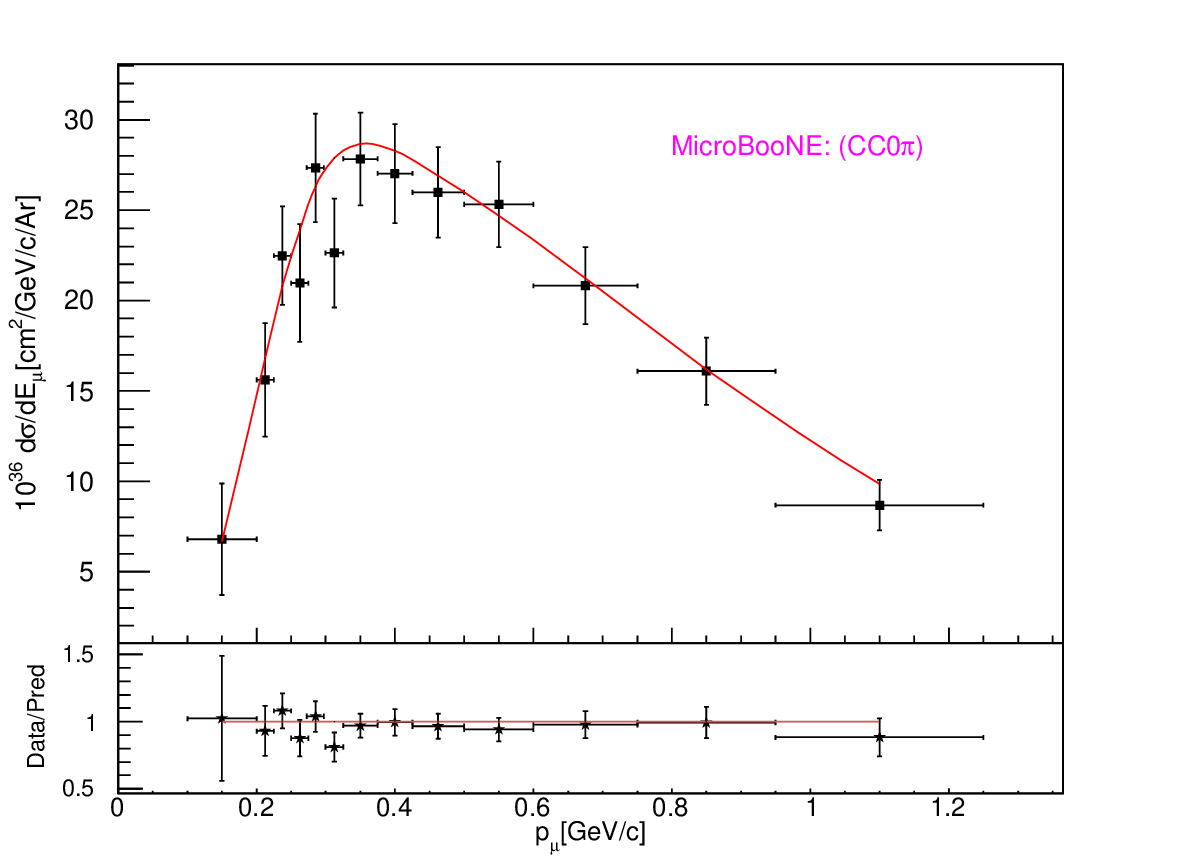}
  \end{center}
  \caption{\label{Fig8}
    Flux-integrated inclusive $d\sigma/dp_{\mu}$ cross section per argon
    nucleus as a function of muon momentum $p_{\mu}$. The bottom subpanels show
    the comparison between the data and the theoretical predictions.     
}
\end{figure*}
Thus, CCQE-like (CCQE + MEC) interactions provide the dominant contribution to
pionless $\nu_{\mu}\text{Ar}$ scattering.

We calculated the flux-integrated double-differential cross section 
$d^2\sigma/dp_{\mu} d\cos\theta$  using 
the muon momentum $p_{\mu}$ and $\cos\theta$ bins from Ref.~\cite{Micro6}.
Figures~\ref{Fig6} and~\ref{Fig7} show measured flux-integrated 
$d^2\sigma/dp_{\mu}d\cos\theta$ cross sections as functions of the muon
scattering angle for several bins of the muon momentum in the range
$0.1\leq p_{\mu}\leq 1.0$~GeV/$c$ as compared with calculations.
Overall, agreement is observed between the calculated and measured cross
sections within experimental uncertainties. However, the small statistics and
large experimental uncertainties do not allow for a definitive conclusion.
The figures show experimental uncertainties, defined as the square root of the
diagonal elements of the extracted covariance matrix which exceed 15\%.
\begin{figure*}
  \begin{center}
    \includegraphics[height=12cm, width=18cm]{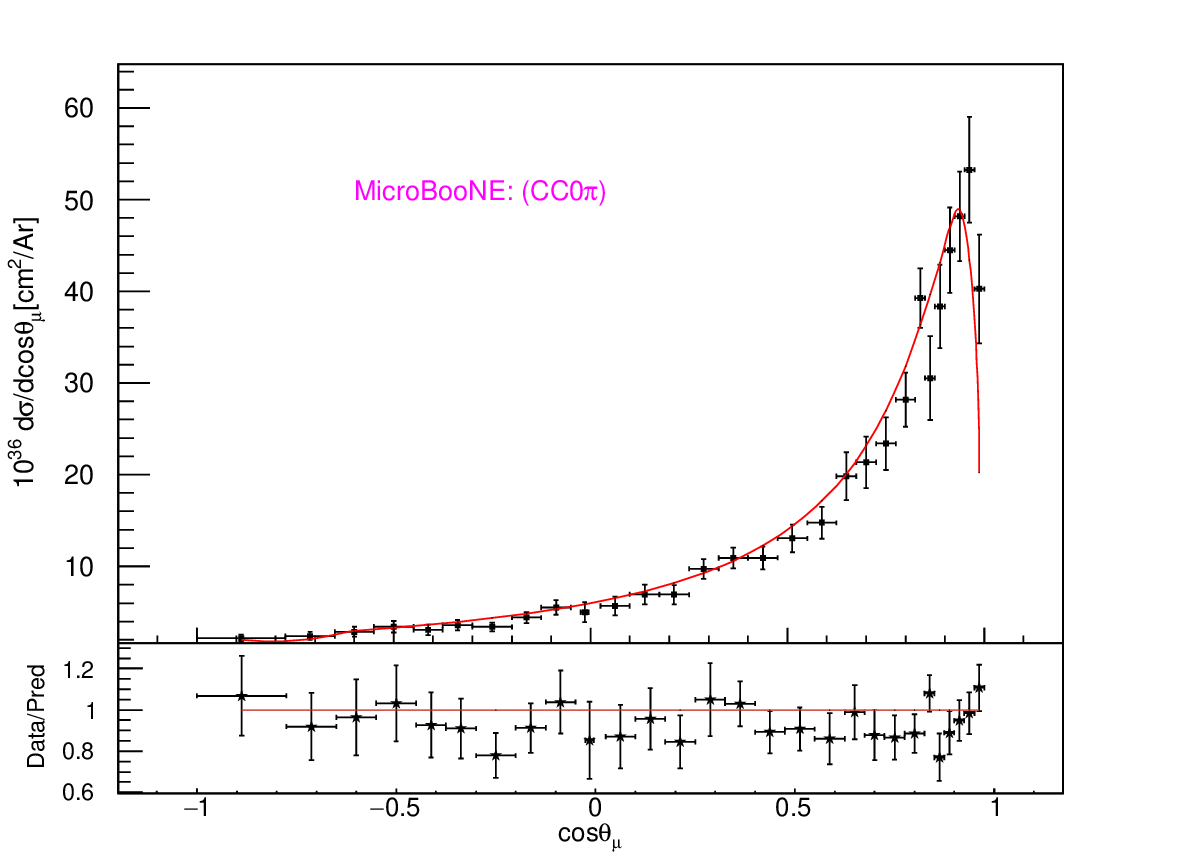}
  \end{center}
  \caption{\label{Fig9}
    Flux-integrated single-differential $d\sigma/d\cos\theta$ cross section per
    argon nucleus as a function of the muon scattering angle $\cos\theta$. The
    bottom subpanels show the comparison between the data and the theoretical
    predictions.     
}
\end{figure*}
The flux-integrated single-differential cross sections of pionless neutrino
scattering on an argon nucleus, $d\sigma/dp_{\mu}$ and $d\sigma/d\cos\theta$
(for $p_{\mu} > 1$ GeV/c), are presented in Figs.~\ref{Fig8} and ~\ref{Fig9},
respectively. The measured $d\sigma/dp_{\mu}$ cross section is shown as a
function of the muon momentum in Fig.~\ref{Fig8}. Figure~\ref{Fig9} shows the
$d\sigma/d\cos\theta$  cross section as
a function of the cosine of the measured the muon scattering angle. The data are
compared to the calculated differential cross sections. There is good agreement
between the calculated and measured cross sections. As can be seen in
Fig.~\ref{Fig9}, the calculated cross section $d\sigma/d\cos\theta$ is in
overall agreement with the data within the experimental uncertainties, except
for the highest $\cos\theta$ bin, where the calculated cross section
underestimates the measured one.

\subsection{Comparison of $d\sigma/dT_{\mu}$ differential cross sections of
  pionless neutrino scattering measured in the MiniBooNE and MicroBooNE
  experiments}

Currently, a large amount of data has been accumulated on electron and neutrino
 scattering on carbon. Therefore, the question arises as to how these data can
be used to test the accuracy of calculations for neutrino-nucleus scattering
cross sections on argon. To achieve this, it is necessary to investigate how
nuclear effects differ between scattering on carbon and argon nuclei. Since
these effects depend on the incoming lepton energy, they must be studied within
 identical energy ranges, i.e., using the same neutrino beam.
This setup naturally requires different detectors, for example, LArTPCs and
scintillation detectors. Consequently, the detection methods and neutrino event
reconstruction algorithms - and therefore particle identification, as well as
the precision of their momentum and energy measurements - will differ, leading
to different experimental uncertainties in the measured cross sections. Thus,
to extract information about the differences in nuclear effects between carbon
and argon scattering from the analysis of measured cross sections, the
experimental uncertainties must be significantly smaller than the potential
differences in the cross sections arising from these nuclear effects.

At different neutrino energies, different nuclear interaction processes
dominate. For instance, at the energies of the BNB flux, CCQE-like
processes dominate (accounting for nucleon Fermi motion and FSI effects),
whereas at DUNE energies, the main contribution comes from resonance
production and deep inelastic scattering (where EMC effects play a major role).
Consequently, a comparison of pionless neutrino scattering cross sections
measured in the MiniBooNE (carbon) and MicroBooNE (argon) experiments, as well
as cross sections to be measured at the DUNE near detector, may allow for the
investigation of differences in nuclear effects across a wide energy range.

At BNB flux energies, CCQE-like processes provide the main contribution to
pionless neutrino-nucleus scattering. Therefore, we compare the
differential cross sections $d\sigma/dT_{\mu}$ measured in the MiniBooNE
~\cite{MiniB} and MicroBooNE experiments as a function of the muon kinetic
energy, scaled to the number of neutrons in carbon and argon.
\begin{figure*}
  \begin{center}
    \includegraphics[height=13cm, width=14cm]{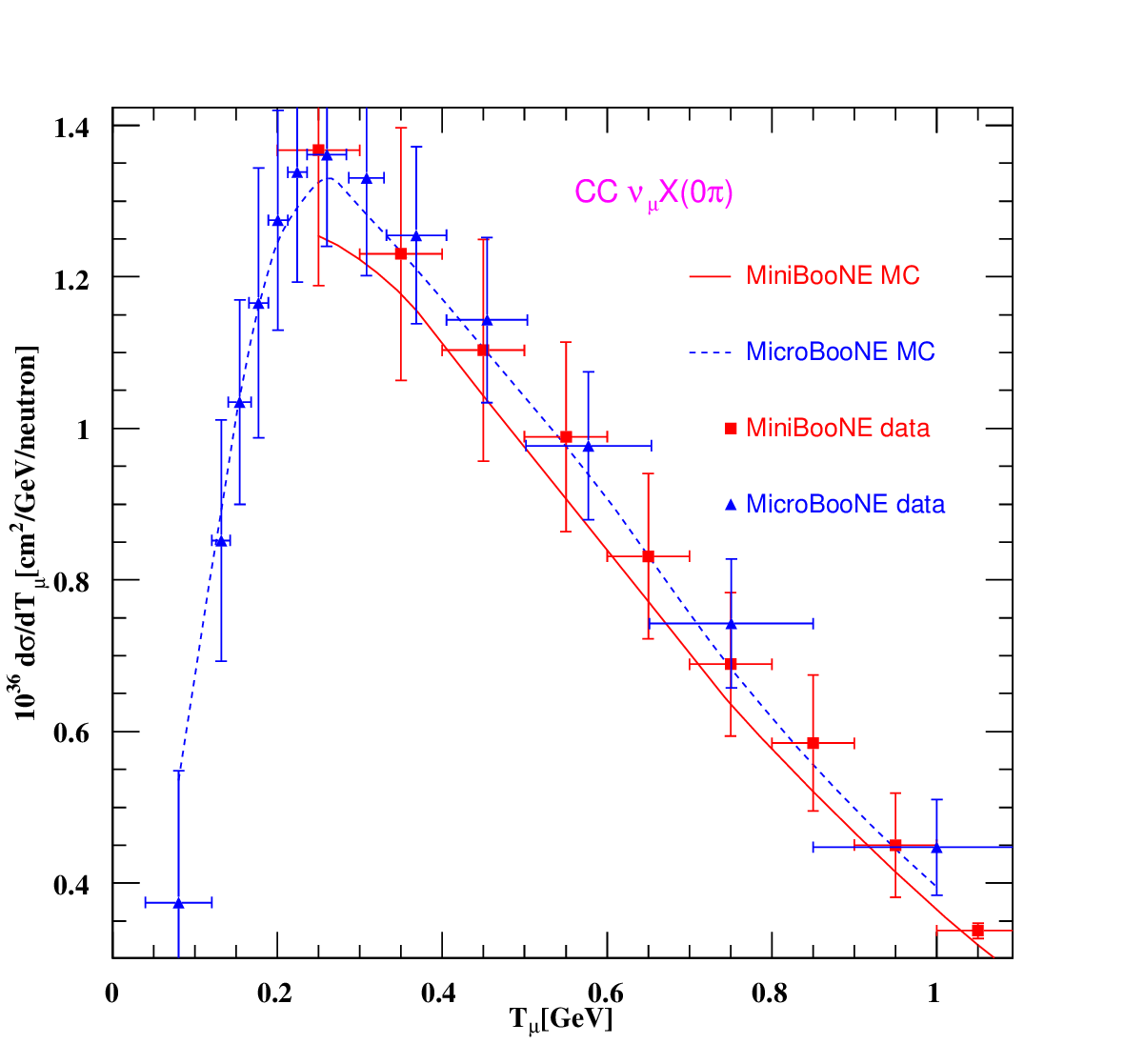}
  \end{center}
  \caption{\label{Fig10}
Measured $d\sigma/dT_{\mu}$ cross sections per neutron for pionless neutrino
scattering on carbon (triangles)~\cite{MiniB} and argon (squares)~\cite{Micro6}
nuclei as a function of the muon kinetic energy. The corresponding
calculated results are displayed as a solid line for MiniBooNE and a
dashed line for MicroBooNE.     
}
\end{figure*}
Additionally, the cross sections $d\sigma/dp_{\mu}$ measured by MicroBooNE
experiment were converted into $d\sigma/dT_{\mu}$.
 This conversion of the differential cross section from momentum $p_{\mu}$
transformation: $d\sigma/dT_{\mu} =(E_{\mu}/p_{\mu}) \cdot d\sigma/dp_{\mu}e$.
Figure 10 shows the measured $d\sigma/dT_{\mu}$ cross sections per neutron for
 pionless neutrino scattering on carbon and argon nuclei as a
 function of the muon kinetic energy, along with the corresponding calculated
 results. The calculation for carbon was obtained from Ref.~\cite{BAV8}. 
 The difference between the calculated cross sections for carbon and argon is
 approximately 6\% and depends weakly on the muon energy. Meanwhile, a
 comparison of the converted MicroBooNE differential cross section with the
 MiniBooNE data demonstrates good agreement within the experimental
 uncertainties. Thus, these uncertainties preclude a quantitative estimation of
 the difference in nuclear effects between the two nuclei.

\section{Conclusions}

In this work, we calculated the flux-inegrated differential cross sections for
inclusive and pionless neutrino-nucleus scattering on argon nuclei and compared
 the results with MicroBooNE data. Both the MicroBooNE
and MiniBooNE experiments utilize the BNB flux at energies where CCQE-like
interactions dominate. We analyzed these data within the RDWIA+MEC approach.
The measured and calculated inclusive cross
sections are in good agreement within experimental uncertainties.
The contribution of the CCQE process increases with muon energy, exceeding 65\%
above $\var_{\mu}=1.2$~GeV. For pionless neutrino scattering, general agreement
between the measured and calculated $d^2\sigma/dp_{\mu}d\cos\theta$ differential
cross sections is observed within the large experimental uncertainties.
Concurrently, the $d\sigma/dp_{\mu}$ and $d\sigma/d\cos\theta$ differential
cross sections show good agreement with the MicroBooNE data.

To investigate the differences in nuclear effects for neutrino scattering on
carbon and argon, we compared the differential cross sections measured in the
MiniBooNE and MicroBooNE experiments, scaled to the number of target
neutrons. Our calculations indicate that this difference is approximately 6\%.
A comparison of the two experimental datasets demonstrates consistency within
their experimental uncertainties, which exceed 10\%. Consequently, to study
the differences in nuclear effects in neutrino scattering off carbon and argon
targets at energies below 2~GeV in future experiments, it is crucial to
measure differential cross sections with an accuracy better than 6\%.
 
\section{Data availability}

The data that support the findings of this article are openly available~\cite
{Micro5, Micro6, MiniB, BAV8}

\section*{Acknowledgments}

The author greatly acknowledges G. Zeller for fruitful discussion the results
obtained in this work.
%

\end{document}